\documentclass[a4paper,fleqn,usenatbib]{mnras}

\usepackage{graphicx}   
\usepackage{amsmath}    
\usepackage{amssymb}    
\usepackage{amsfonts} 

\usepackage{newtxtext,newtxmath}

\title{Planet-planet scattering in systems of multiple planets of unequal mass.}

\author[F. Marzari]{
F. Marzari,$^{1}$
\\
$^{1}$DFA, University of Padova, Italy \\}

\begin{document}
\maketitle

\begin{abstract}
A large sample of planet-planet scattering events for three planet systems with different orbital separations and masses is analyzed with a multiple regression model. 
The dependence of the time for the onset of instability on the masses of the planets and on their initial orbital separations is modeled with a quadratic function. The same analysis is applied to the timespan of the chaotic evolution dominated by mutual close encounters. The configurations with the less massive planet on an outside orbit are stable over longer timescales. The same configuration leads to shorter chaotic evolution times before the ejection of one planet. In about 70\% of the cases the lighter planet is the one escaping from the system.  If a different separation is assumed between the inner and outer planet pairs, then the dominant effect on the instability time is due to the pair with the smaller separation, as a first approximation. 

\end{abstract}


\section{Introduction}
\label{Introduction}

Planet-planet scattering is a dynamical process where the mutual gravitational interactions between the planets in multiple planetary systems cause significant changes in their orbits. This typically leads to the ejection of one or more planets from the system, which become rogue planets, collisions between planets or with the central star and dramatic variations of their orbital parameters (\cite{weimar1996,rasioford1996,chambers1996,lin1997,marzari2002, marza2005ApJ, Chatterjee2008,raymond2009, Raymond_2010, Nagasawa2011,petrovich2014, mustill2014,petrovich_rafikov2014,deienno2018,raymond_barnes2008,rice2016,rice2018} see also \cite{davies2014} for a review). 

A system with multiple planets which is stable during the planetary growth within the circumstellar disc may become unstable due to different mechanisms like: 1) convergent migration
\citep{masset2001,Leepeale2002}, caused either by different planet masses and then different migration speeds or due to a slowing down of the inner planet inward drift when close to the star because of the formation of an inner disc hole
2) the dissipation of the gaseous disc and the termination of its damping effects on the planet orbits \citep{moor2005,matsu2008,pichi2023}, 3) interaction with the residual planetesimal population (see for example \cite{gomes2004,ghosh2023}) and 4) interaction with a passing star in a cluster  \citep{mal2007,mal2009,mal2011,zaka2004} possibly occurring after the dissipation of the disc which might be able to damp the planetary eccentricities after the flyby \citep{marpi2013M,marpi2014}. Once the system becomes unstable, the eccentricities grow and the close encounters due to orbital crossing result in significant energy and angular momentum exchanges.
The chaotic phase ends when one or more planets gain enough energy to escape the gravitational pull of the star, becoming rogue planets, or when two planets collide forming a bigger body or after a planet impacts the star. The surviving planets have their orbits significantly different from the original ones and are usually pushed into highly elliptical or inclined orbits. 

P-P scattering explains the large eccentricity values observed among exoplanets while 
the interplay between P-P scattering and tides may lead to the formation of hot/warm planets. In addition, many transiting multiple planetary systems appear to be on the verge of instability with lifetimes of the same order of their age \citep{fang2013,pu2015,volk2015,obertas2023}. It is possible that a significant
fraction of these systems underwent dynamical instabilities
 in the past and the configuration we observe at present is
a consequence of the temporary chaotic evolution.
 Collisions, ejections and mergers occurring in this phase
would lead to a new orbital configuration with fewer planets
and wider orbital spacings. The new architectures may either
be stable or be characterized by longer instability times.
This process may repeat leading each time to new systems with fewer planets on wider orbital spacings. This scenario is supported by recent observations
\citep{poon2020, lammers2023,ghosh2024} suggesting that giant impacts can reproduce some trends observed in the exoplanet population.  Comparing the orbital architecture of young systems with aged ones would shed new light on the  extent of the rearrangement that transiting multi-planet systems may undergo during their evolution.

The possibility of comparing old and young multi--planet systems to estimate the amount of dynamical heating and potential periods of instability,  for example using the
Angular Momentum Deficit (AMD) index  \citep{laskar2017}, is an important step to comprehend the evolution of multiplanet systems to their final stage.  Therefore, understanding the P-P scattering mechanism is crucial for models of planet formation and the long-term evolution of planetary systems. 

All previous numerical modelings for estimating the time of the onset of instability $t_i$ in a multi--planet systems are based on N--body numerical simulations where the initial planets have the same mass. This assumption is made in order to reduce the dimensionality of the problem and $t_i$ is evaluated against the initial planetary separations measured in mutual Hill's spheres \citep{chambers1996, marzari2002,Chatterjee2008,rice2018}.   
While the dynamics can be scaled with the keplerian frequency for different values of the semi--major axis of the planets (apart from the number of collisions, see Table 1 of \cite{marzari_nagasawa2019}), this cannot be done for the initial mass distribution.
In this paper I explore a model where also the initial masses of the planets are varied. A multiple regression is performed in order to obtain a fitting formula which depends on $m_1,m_2,m_3$, the masses of the three planets, and $K$ measuring the initial spacings as a function of the mutual Hill's sphere. 
The use of a single spacing $K$ has been adopted in previous studies to typically model planetary systems which emerge from the circumstellar discs after a quasi--static growth. However, if migration occurs, different separation may occur between the planets after the dissipation of the disc. Also this case is discussed.

In Sect. \ref{parameters} the important parameters of the problem are discussed and it is outlined the way in which the N--body numerical integrations are set up. 
In Sect. \ref{model} it is introduced the regression model used to analyze the outcome of the N--body simulations.  In Sect. \ref{outcome1} the output is discussed while in Sect. \ref{limits} the limits of the interpolation process are commented. Sect. \ref{masdep1} is devoted to the derivation of some physical results concerning the relation between the mass configuration and the time interval for the onset of chaos. In Sect. \ref{masdep2} the length of the chaotic phase is interpolated for different values of the parameters showing that systems where the less massive planet is the outermost have a shorter chaotic evolution. 
In Sect. \ref{correlation} the statistics of the escaping planets vs. their mass is given while in Sect \ref{doubleK} the case with different values of $K$ between the inner and outer planet pairs is discussed. 
Finally, in Sect \ref{discussion} the results are commented. 

\section{Relevant parameters of the P-P scattering dynamics}
\label{parameters}

The timescale for the onset of instability and chaos $t_i$ in 3--planet systems is dictated by the initial semi--major axis of the planets and by their masses (possibly scaled to the stellar mass). Since they form in a circumstellar disc, it is reasonable to assume that their initial eccentricity and inclination are low and, in the forthcoming modeling, both these orbital elements are set to a random value smaller than 0.001. Therefore, the dynamics is determined by six free parameters, the three planetary masses and the three initial semi--major axes of the planets. To reduce the number of free parameters, instead of using the initial semi--major axes, the planet separations are  expressed in terms of $K$ times their mutual  Hill's sphere (the scaling with the radial distance of the inner planet from the star is expected to be Keplerian). The mutual Hill's sphere of radius $R_H$  is defined as 

\begin{equation}
\label{eq:rh}
R_H = 
\left(\frac{m_i+m_{i+1}}{3 M_0}\right)^{1/3} \frac {(a_i+a_{i+1})} {2} , 
\quad i=1,2,\ldots,
\end{equation}

where $M_0$ is the mass of the central star,  $m_i$ are the masses of the planets and $a_i$ their initial semi--major axis. Since $R_H$ depends on both $a_i$ and $a_{i+1}$, the value of $a_1$ is initially fixed and an iterative procedure is used to compute $a_2 = a_1 + K a_1$ where at each iteration  both $R_H$ and $a_2$ are updated. This procedure converges very quickly and it gives the first pair of semi--major axes $a_1$ and $a_2$ for a given value of $K$. The same algorithm is then used to compute $a_3$ from $a_2$. This method allows a detailed computation of the mutual $R_H$ for different values of $K$ but it is different from the method adopted for example in \cite{marzari2000} and \cite{chambers1996} where the value of $a_2$ is typically set equal to $a_1$ while in others it is not specified \citep{Chatterjee2008,rice2018}. This may introduce differences in the plots of the instability times vs. $K$ shown here compared to those given in other papers. The initial value of $a_1$ is set to 5 au in all the numerical simulations.

Thanks to the choice of $K$ to parametrize the initial separation between the planets, our model now depends only on four parameters: $m_1$, $m_2$, $m_3$, $K$. 

Part of the planet mass dependence of $t_i$ is included in the parametrization of $t_i$ on $K$ through  the definition of the Hill's sphere. However, this parametrization does not comprise 
in full the dependence of $t_i$ on $m_1, m_2$ and $m_3$. This was already observed in \cite{marzari2014} when studying the stability properties of 2 and 3 planets with FMA (Frequency Map Analysis, see for example \cite{laskar1992,laskar1993}), a chaos indicator, and it is confirmed by Fig. \ref{fig:uno} where $t_i$ 
and $t_c$ are plotted vs. $K$ for different mass
combinations [$m_1$, $m_2$,$m_3$]. 
The values of $t_i$ depend on the initial value $a_i$, however it is expected that they scale with the keplerian period i.e. with 
$\sqrt(a_1^3)$. 

\section{The numerical model}
\label{model}

The goal of the numerical model is to perform a multiple regression using as dependent variable the time of the onset of instability $t_i$ and, as independent variables, $m_1$, $m_2$, $m_3$ and $K$.  The timespan of the chaotic evolution $t_c$ is also computed as the interval of time from the first close approach between the planets up to the ejection of one of them or a collision either between two planets or of a planet with the star. Both $t_i$ and $t_c$ (given in years) are fitted on the basis of the independent variables. The individual steps of the modeling are the following.

\begin{itemize}
\item A large number of numerical integrations is performed with three planets around a solar mass star by using the RADAU integrator \citep{radau1985} well suited to handle close approaches between the planets. The maximum integration timespan is 10 Gyr but the run is halted whenever a planet escapes on a hyperbolic orbit. If $t_i$ exceeds 10 Gyr the run is halted and the outcome is not enlisted in the final file. An initial set of three random masses is selected for the planets in the range from 0.01 to 1.5 $M_J$. Once the masses are chosen, the semi--major axes are computed with $a_1 =5$ au and $a_2$ and $a_3$ derived for different values of $K$ ranging from 2 to 7 with an increasing step of 0.05. Being the most important parameter, it is over--sampled respect to the masses of the planets. At the end of each run the masses of the planets, the value of $K$, $t_i$, $t_c$ and the index of the escaping planet or of that impacting on the star are written in a file. These runs are repeated for different mass combinations until a total sample of 10000 P-P scattering cases are accumulated. 

\item To perform the linear regression I have used the open-source machine learning library for Python Scikit-learn \citep{pedre2011}. A polynomial regression of second degree has been performed to interpolate the values of $log_{10}(t_i)$ and $log_{10}(t_c)$ as a function of $m_1$, $m_2$, $m_3$, $K$. A quadratic interpolation has proven to give a lower value of the mean squared error of the regression. The functional form of the fitting function and the derived coefficients are given in appendix. It is noteworthy that this approach is not a machine learning one since the parameters to be used are fixed from the beginning by the physics of the dynamical system.  
\end{itemize}

The linear regression is performed on numerical simulations which end at 10 Gyr. Therefore, it is expected that the validity of the polynomial regression is limited to this interval of time. Beyond 10 Gyr, the regression may catch the trend, thanks also to the fine sampling of the $K$ parameter, but it is not granted that it is correctly modeling the evolution of the instability time. However, 10 Gyr is the average lifetime of solar--type stars and the vast majority of multi--planet systems discovered so far have ages in this range and our model can be safely applied to them. 

Hereinafter, to indicate the three values of mass of the planets  the symbolic writing  [$x_1$,$x_2$,$x_3$] will be used where 
$m_1 = x_1 \cdot m_J$ and so on. 

\section{Outcome of the numerical simulations} 
\label{outcome1}

\begin{figure}
\hspace{-1.5cm}
   \includegraphics[width = 80 mm,angle=-90]{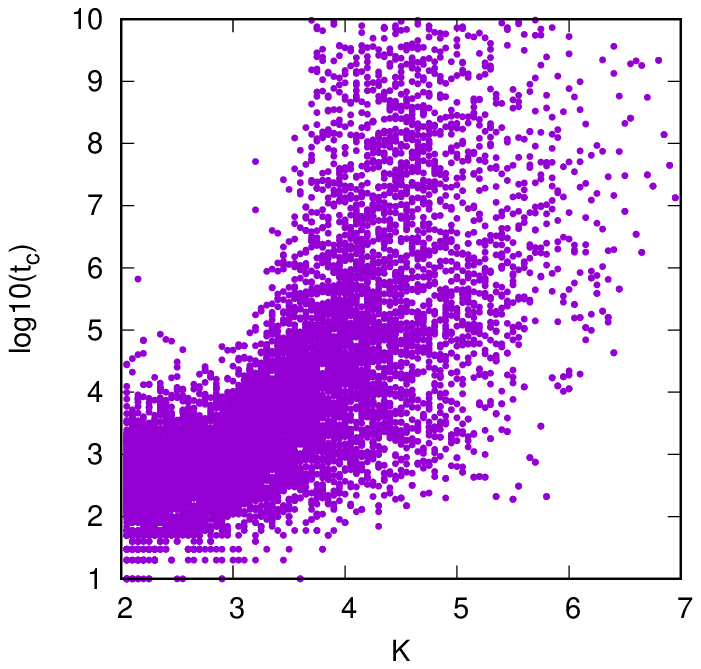}
   
\hspace{-1.5cm}
   \includegraphics[width = 80 mm,angle=-90]{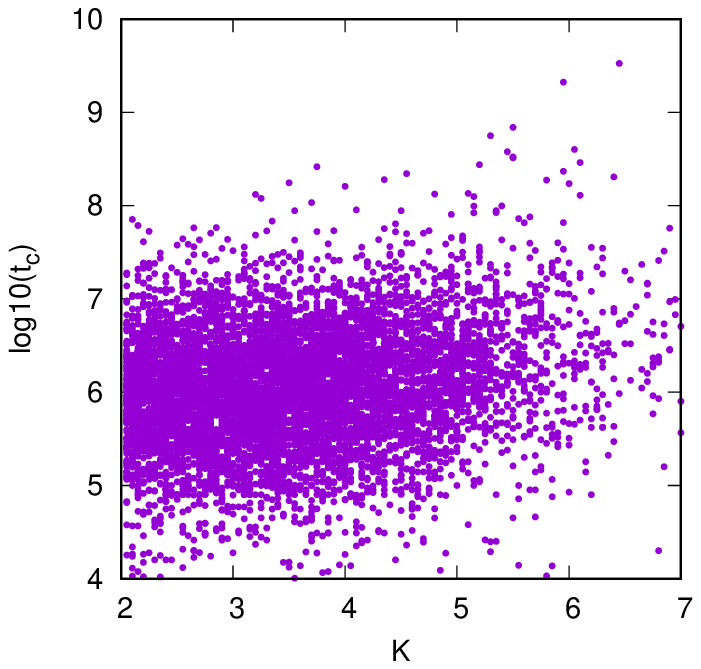}
   
      \caption{ Outcome of the 10000 numerical integrations for different values of $m_1, m_2, m3, K$. On the top panel the 
      $log_{10}(t_i)$ is shown as a function of $K$ while on the bottom panel $log_{10}(t_c)$ is shown. The aligned values on the top panel are related to the choice of the output timestep which was set to $10$ years.}
         \label{fig:uno}
   \end{figure}

The outcome of all the numerical integrations performed to compute the quadratic regression model are shown in 
Fig. \ref{fig:uno}. Superimposed to the natural increasing trend of $log_{10}(t_i)$ with $K$ there is an overall diffusion of the data  due to both the natural chaotic nature of the dynamics and the mass dependence. What is estimated by the data fit is a time average for the instability timescale and, for fixed values of the planet masses, the observed instability times cluster around the fitting curve, as shown in  Fig. \ref{fig:bad}. The quadratic fit can be interpreted as a Lyapunov time giving an estimated timespan before the onset of the chaotic evolution, timespan that grows quickly for larger values of $K$. The computed $t_i$s evolve along this curve with a dispersion due to their chaotic nature. In addition, there is also the dependence on the planet masses which further diffuse the data since smaller masses lead to longer values of $log_{10}(t_i)$ compared to bigger masses. On the bottom panel, the values of $t_c$ are shown. There is an overall increasing trend of $log_{10}(t_c)$ with $K$ suggesting that a larger initial separation between the planets leads to a more extended period of instability dominated by close encounters before the ejection of a planet or collision. Even in this case, there is a significant dependence of $t_c$ on the planet masses, in addition to the natural chaotic diffusion, which can be disentangled only with the multiple regression model.  

\section{Limits of the approach}
\label{limits}

\begin{figure}
\hspace{-1.8cm}
   \includegraphics[width = 80 mm,angle=-90]{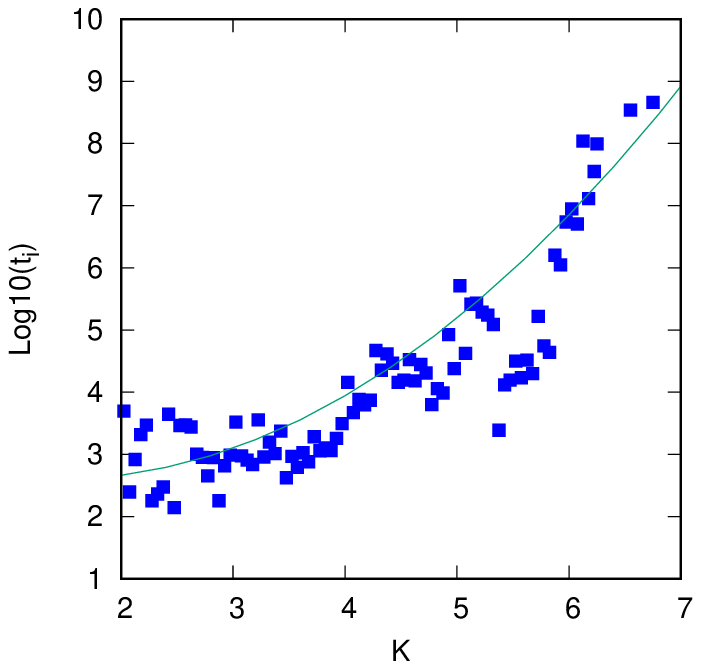}
   
\hspace{-1.8cm}
   \includegraphics[width = 80 mm,angle=-90]{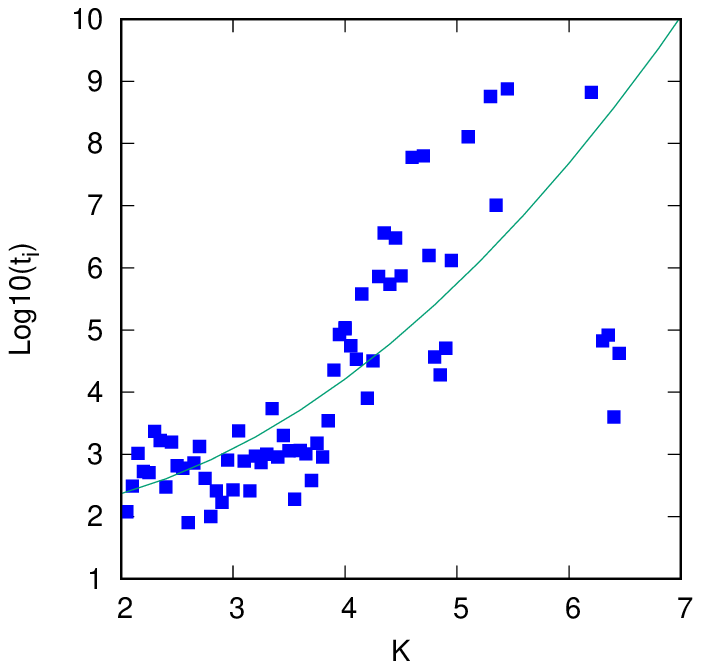}
   
      \caption{ Comparison between the data from direct N--body numerical integrations and the multi regression fit model. On the upper panel a case with a mass configuration [0.01,0.28,0.54] is shown. The fit is good and it well reproduces the growing trend of $t_i$ with $K$. On the lower panel a different mass configuration given by [0.82,0.23,0.97] is shown. After an initial growth of $t_i$, followed by the fit, the instability times decreases again for larger values of $K$ due to resonant interactions.}
         \label{fig:bad}
   \end{figure}

As already discussed in \citep{marzari2002}, resonances between the planets may affect the timescale required for the system to become unstable. In this case, the increase of $t_i$ with $K$ is not well approximated by a quadratic function but the trend is more complex and cannot be well fitted by a low degree polynomial. In Fig. \ref{fig:bad} we compare the outcome of the direct N--body numerical integration with the quadratic fit of $t_i$ in two different mass configurations. On the upper panel the fitting curve matches the data up to long timescales and the fitting formula is predictive of the timescale for the onset of instability of the system. On the other hand, on the lower panel, the fit is less robust since mean motion resonances cause a reduction of the timescale beyond $K \sim 6$, a phenomenon that the quadratic fit is not able to reproduce. In this case a more complex approach is needed, where the location of the 2--body, 3--body and secular resonances are mapped to look for superposition 
and chaos. Such an approach requires an ad--hoc modeling of each individual system and it is beyond the scope of this paper. 

The behaviour shown in Fig. \ref{fig:bad} reduces the predictive power of the regression model when applied to a single mass configuration but it can still describe the general trend of $t_i$ as a function of different mass configurations where the individual behaviour is less important. This is discussed in the next Section. 

 It is also impossible to estimate an average dispersion around each individual fit obtained with the regression model (see for example Fig. \ref{fig:bad}) since the fit is performed on 4 parameters and it is not possible to evaluate the dispersion due to a single parameter for a fixed value of each of the three mass values. A possibility would be to perform an additional regression on the difference between each numerical timescale, for a given set of $[K,m_1,m_2,m_3]$, and the predicted value. This would give a field of dispersion, and the possibility of computing an average error for a given set of mass values for each $K$ value and not for the curve fitting $t_i$ vs. $K$. However, it looks a complex procedure and it would not be useful to predict the trends as a function of the masses which is the most interesting aspect of the regression model. 

\section{Mass dependence of the instability time} 
\label{masdep1}

\begin{figure}
\hspace{-1.8cm}
   \includegraphics[width = 80 mm,angle=-90]{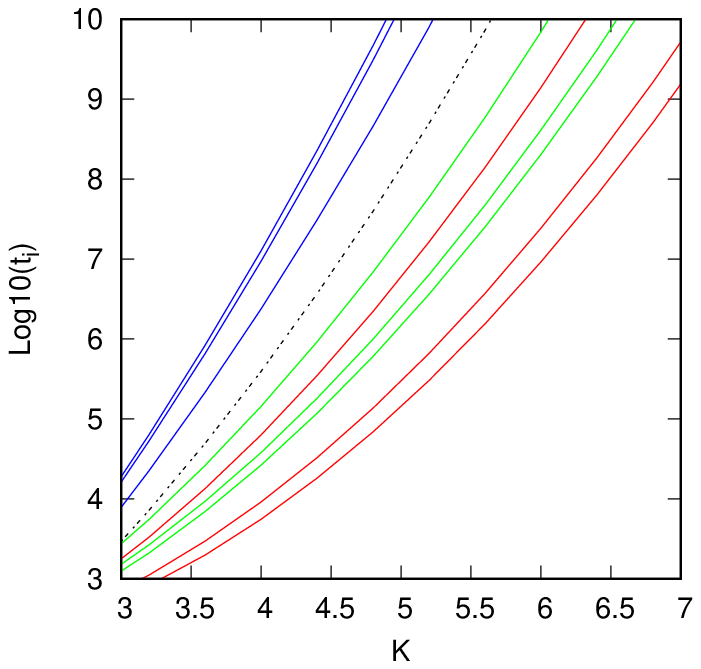}
   
\hspace{-1.8cm}
   \includegraphics[width = 80 mm,angle=-90]{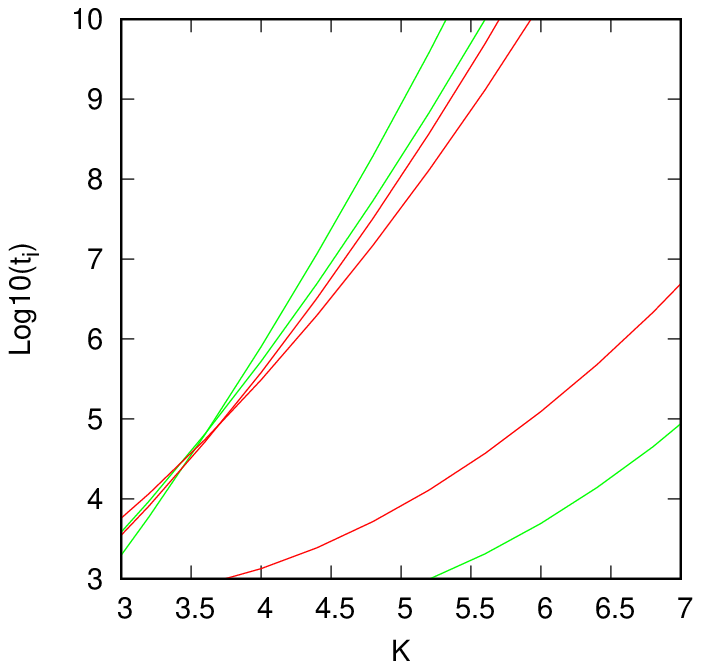}
   
      \caption{Dependence of $t_i$ on $K$ for different mass combinations of the three planets. On the upper panel the green lines are for the following mass combinations: [0.01,1,1] (bottom line), [0.1,1,1] and [0.5,1,1] (top line). The red lines are for [1,0.01,1] (bottom lime), [1,0.1,1] and [1,0.5,1] (top line). Finally, the blue lines are for [1,1,0.01] (top line), [1,1,0.1] and [1,1,0.5] (bottom line). The black dashed line corresponds to the classical reference case with equal mass planets i.e. [1,1,1].  On the bottom panel, the following mass combinations are shown: with green lines [1.5,0.1,0.1], [0.1,1.5,0.1] (top line), and [0.1,0.1,1.5] (bottom line), red  lines are for 
      [1,0.1,0.1], [0.1,1,0.1] (top line) and [0.1,0.1,1] (bottom line).
}
         \label{fig:due}
   \end{figure}

Using the linear regression model developed from the test data sample, we can predict the dependence of $t_i$ vs. $K$ for different mass combinations of the planets. It is possible to outline changes in $t_i$ when the mass of the inner or outer 
planet is larger or smaller reproducing the different combinations observed in multi--planet exoplanetary systems. In
Fig. \ref{fig:due} top panel, the mass of one planet is changed from the lower value considered here (0.01 $m_J$) to 1 $m_J$ and this is done for the inner, medium and outer planet, respectively. The fitting curves show an increase in 
$t_i$ when the mass of the smaller planet grows from 0.01 to 1 $m_J$ in the configurations where the less massive planet is in the inner or intermediate orbit. This indicates the tendency of these systems towards longer $t_i$ and the dashed curve representing the 
configuration [1,1,1] is a sort of limiting curve. However, if the less massive planet is the outer one, then the $t_i$ 
is remarkably longer (blue lines) for increasingly smaller values of $m_1$. This suggests that multiple planetary systems with the less massive planet on an outside orbit are more stable. 

On the bottom panel of Fig. \ref{fig:due} I consider two small equal mass planets (0.1 $m_J$) and a more massive planet on an inner, intermediate and outer orbit, respectively. The green lines show the case where the heavier planet has a mass equal to $m=1.5 m_J$ while the red lines are for
$m=1.0 m_J$. When the outer planet is the massive one, the values of $t_i$ are extremely short showing that this configuration is highly unstable.
As a consequence, a Jupiter size planet on an external orbit
may easily destabilize a system of two planets internal to its orbit.
On the other hand, the configurations with a massive planet orbiting
in between two less massive planets (top green and red lines in Fig.
3 bottom panel) or inside them are significantly more stable.

\section{Dependence of the chaotic evolution time--span on the regression parameters}
\label{masdep2}

The highly chaotic evolution of the planets begins after the first close encounter between a planet pair. From then on all three planets have mutual encounters and the changes in their orbital elements following each close approach are impulsive and strong until one or more planets are ejected from the system or impact with the star or between themselves. 
 Practically in all cases  it is expected that the chaotic phase ends when one planet is ejected from the system, or a collision occurs. The two surviving planets are typically left in a stable dynamical configuration (see for example \cite{marzari2002, marza2005ApJ, Chatterjee2008,Nagasawa2011,petrovich2014}). 
I define $t_c$ as the time interval of the highly chaotic evolution dominated by close approaches between the planets, from the first close encounter to the ejection/collision of one planet. The value of $t_c$ is computed for all the different systems in our sample and the outcome is shown in the bottom panel of Fig.\ref{fig:uno}. 
 Only in a tiny minority of cases the two surviving planets are still in a chaotic state so it takes some additional time to reach the final stable configuration with only one planet. We underestimate the value of $t_c$ in these cases which, however, for a system of initially three planets are not statistically significant.

\begin{figure}
\hspace{-1.8cm}
   \includegraphics[width = 80 mm,angle=-90]{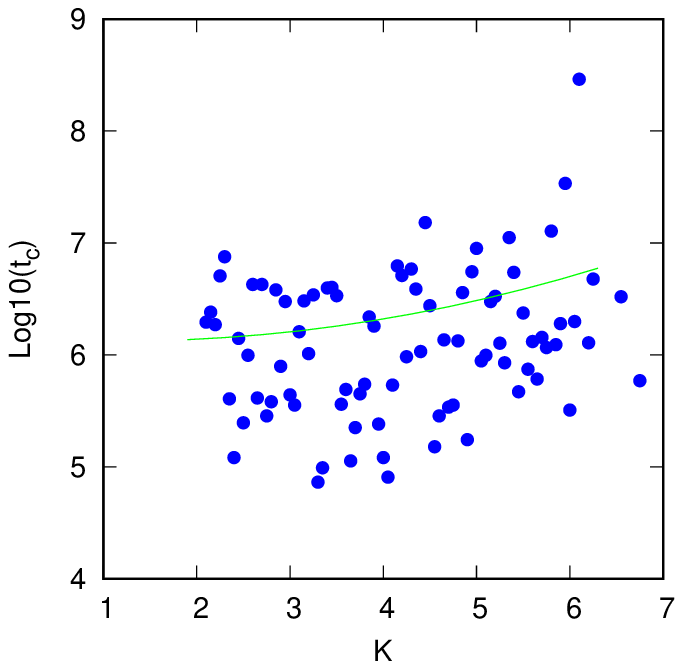}
   
\hspace{-1.8cm}
   \includegraphics[width = 80 mm,angle=-90]{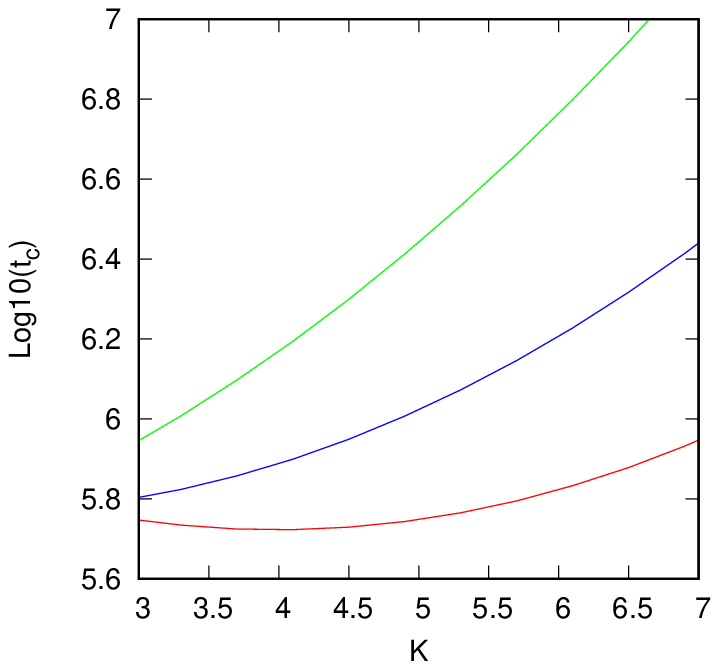}
   
      \caption{In the top panel the
      outcome of the numerical integration (blue filled circles) and the regression fit (green line) are compared in the case with the configuration [0.02,0.28,0.54]. On the bottom panel the predictions of the regression model for the the configurations  [0.1,1,1] (green line), [1,0.1,1] (blue line) and [1,1,0.1] (red line) are compared. The dynamical configuration with the smaller planet inside appears to be the one leading to longer chaotic evolutions. 
}
         \label{fig:settimo}
   \end{figure}

 According to the bottom panel of Fig.\ref{fig:uno}, the rough data suggest a dependence of $t_c$ on the initial planet separation measured by $K$. A regression model, similar to that used to interpolate $t_i$, is applied to the data in order to obtain suitable fitting parameters which describe the mass dependence of $t_c$. In the top panel of Fig. \ref{fig:settimo} the quadratic interpolation is compared to the numerical integration data in the case [0.02,0.28,0.54] confirming and increase in the time--span of the chaotic evolution with $K$. Using the fitting model, it can be shown that $t_c$ is longer for a configuration in which the less massive planet is the innermost. This is illustrated in 
Fig. \ref{fig:settimo} bottom panel where we compare the predictions for $t_c$, as a function of $K$, for three different mass configurations. In the case with [0.1,1,1] (green line) the length of the chaotic evolution is significantly longer respect to the mass configuration [1,0.1,1] (blue line) and [1,1,0.1] (red line). The latter is the configuration leading to the shortest chaotic evolutions with the outer light planet being the one ejected out of the system most of the times. 

\section{Correlation between the initial configuration 
of the planets and the mass of the escaping planet}
\label{correlation}

\begin{figure}
\hspace{-1.8cm}
   \includegraphics[width = 80 mm,angle=-90]{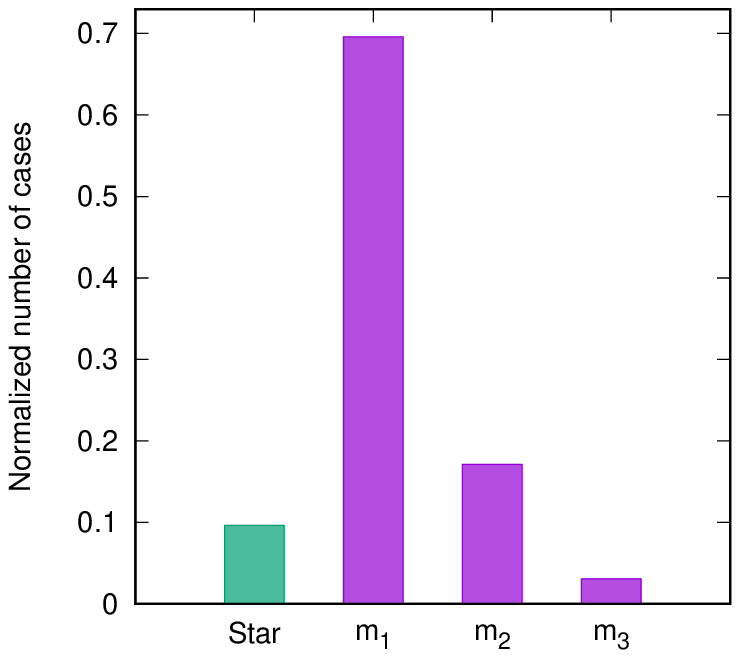}
   
      \caption{Histogram showing the link between the mass of the escaping planet and the initial planet mass distribution. The label $m_1$ marks the fraction of cases where the smallest planet escapes at the end of the chaotic phase, $m_2$ when it is the planet with intermediate mass to escape and $m_3$ when it is the most massive planet to be ejected from the system. The label $Star$ indicates the fraction of cases where one planet impacts on the star.   
}
         \label{fig:histo}
   \end{figure}

At the end of each numerical N--body simulation the mass of the escaping planet is recorded. Therefore, it is possible to statistically infer which is the planet 
more frequently ejected out of the system. The histogram in Fig. \ref{fig:histo} gives the percentages of the escaping planets in terms of their mass relative to the other two. In most cases (about 70\% of cases) the less massive planet is the one ejected out of the 
system after the chaotic evolution. The more massive one is ejected only in less than 2\% of cases. In about 9\% of cases one of the planets impacts the star. 

The statistics somehow confirms the expectations since the less massive planet has  its orbit more easily altered during an encounter with a more massive body. 

\section{Different separation between the planets: two values of $K$}
\label{doubleK}

 In the case of three planets, a single value of $K$, determining the separation of the bodies in terms of Hill's spheres, has been adopted in all previous studies \citep{weimar1996,chambers1996,rice2018}. It is mostly based on the hypothesis that the feeding zone of each growing planet is limited by the mutual Hill's sphere. Under this assumption, the three planets emerge from the circumstellar disc with a separation which is approximately equal in terms of $R_H$ in the case of planet growth in a quiet environment.  In this scenario, the planet--planet scattering may occur at significantly later times respect to the time of the disc dissipation. 

However, planet migration by interaction with the disc may change the initial configuration and lead to a different separation, in terms of $R_H$, between the first and second pair of planets. In this case, two independent values, $K_1$ measuring the separation between planet 1 and  2 (from inside out) and $K_2$ measuring the separation between planet 2 and 3, is preferred to describe the long--term dynamics of the system. 

In the attempt to perform a fit for different values we computed approximately $10^5$ cases with random values of $m_1, m_2, m_3, K_1, K_2$ and tried to interpolate the outcome using these five independent variables. Unfortunately, the fit does not work because of the truncation of  $t_i$ with the smaller of the two values between $K_1$ and $K_2$. 
Different models have been tested but the change in the functional dependence of either of $K$ with the value of the second $K$ makes the regression unreliable.  It is possible that the introduction of additional non physical parameters might allow an AI model to reproduce the behaviour even if the task appears difficult and beyond the scope of this paper.  

\begin{figure}
\hspace{-0.98cm}
   \includegraphics[width = 80 mm,angle=-90]{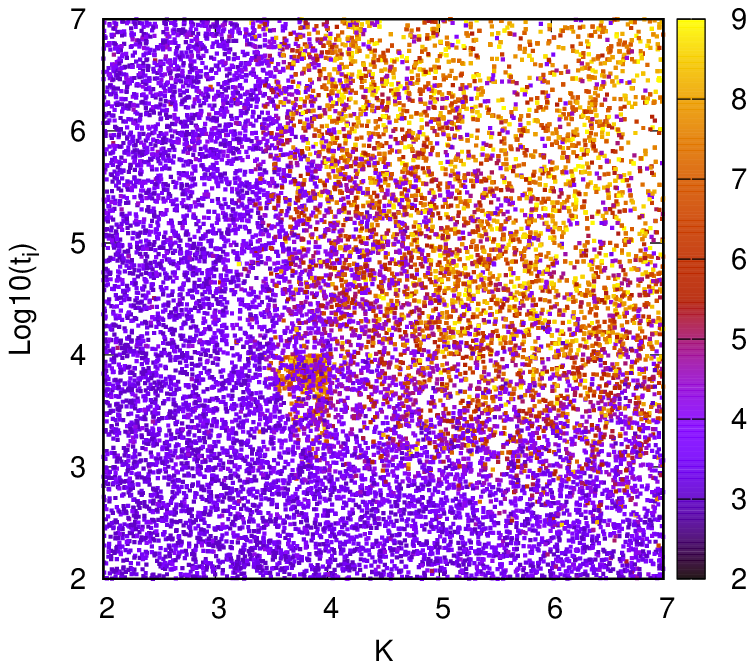}
   
\hspace{-0.98cm}
   \includegraphics[width = 80 mm,angle=-90]{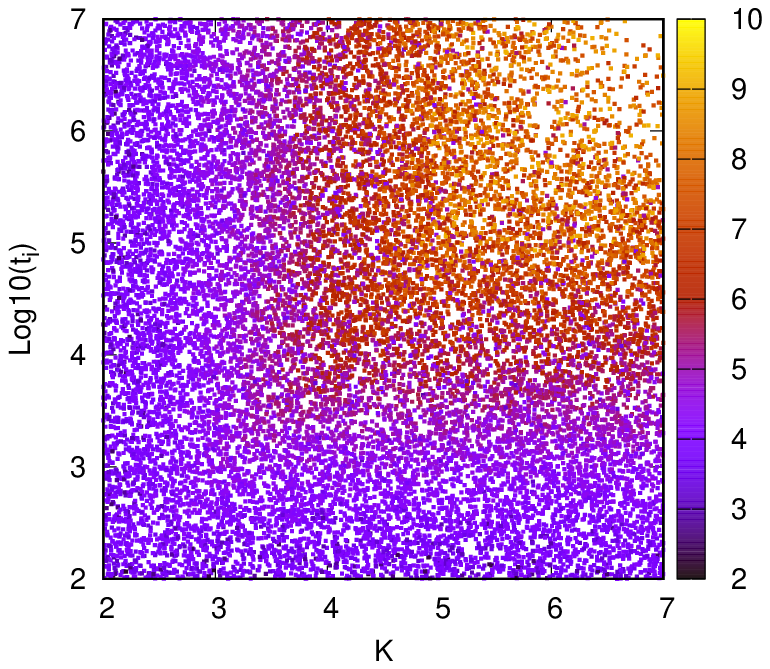}
   
      \caption{In the upper panel the value of $t_i$ is shown for random values of $m_i$, $K_1$ and $K_2$. The different colors code the value of  $log(t_i)$. On the bottom panel a synthetic plot is produced by using the regression parameters of table A1. The values of $m_i$, $K_1$ and $K_2$ are randomly selected and, to compute $t_i$ from the regression model, the smaller between $K_1$ and $K_2$ is assumed as value for $K$.
}
         \label{fig:doppiok}
   \end{figure}

In alternative, the hypothesis that the lifetime before instability is governed by the smaller of the two values of $K_{1}$ and $K_{1}$ has been tested. The results appear promising as shown in Fig \ref{fig:doppiok}. On the upper panel the values of $t_i$ for random values of $m_i$, $K_1$ and $K_2$ are shown while on the bottom panel a synthetic plot is produced by computing the instability time with the fit of Table A1 taking as value of $t_i$ that given by the value of $K$  corresponding to the smaller between $K_1$ and $K_2$. 
Even in this case random value of $m_i$, $K_i$ are adopted and for each pair $[K_1,K_2]$ the smaller between the two is set equal to $K$ in the fitting formula. The two panels appear compatible even if in the data, outcome of the direct numerical integration, mean motion resonances may introduce some features which cannot be reproduced by the regression.  In addition, there is a chaotic noise around the mean value of $t_i$ which in the synthetic model is not present (see Fig.\ref{fig:uno}). This is clearly a rough approach which, however, reproduces surprisingly well the main features of the numerical simulations.  Further study is needed beyond this point in order to understand how the functional dependence of  $t_i$ on $K_1$ depends on the value of $k_2$ and vice--versa.

\section{Conclusions}
\label{discussion}

The main focus of this paper is on the determination of the timescale for the onset of instability and of the length of the chaotic phase as a function of the masses 
of the three initial planets. In order to obtain this goal, a large number of N--body numerical simulations of P-P scattering events for three 
initial planets have been performed with random values of the initial separation between the planets and their masses. The sample of data are then analyzed with a multiple regression model with a quadratic function interpolating the logarithm of the timescales $t_i$ and $t_c$ as a function of $m_1, m_2, m_3, K$. The resulting fitting formula may not be very robust for the individual cases but it can be successfully used to estimate trends 
in both $t_i$ and $t_c$ for varying values of the planetary masses. As stated in Section 2, the dependence of $t_i$ and $t_c$ is not exhausted by parametrizing the two timescales with $K$ times the Hill's sphere. The data show an additional and considerable dependence on the 
planet masses which is accounted for by the multiple regression model. It must also be taken into account that the values of both $t_i$ and $t_c$ can be scaled with the Keplerian period of the inner planet.  This is not true for the total number of collisions. While the outcome of Fig. \ref{fig:histo} holds in term of relative impact frequency between the different bodies in the system, the total number of collisions grows for smaller values of the $a_i$. This because the size of the bodies becomes relevant, for small semi--major axes, compared to the space available for motion. This is illustrated in Table 1 of \cite{marzari_nagasawa2019} with the number of collisions growing for smaller values of the initial $a_i$.

The fitting model predicts that multiple three planet systems with the less massive body on the outmost orbit are more stable. On the other hand, if the outmost planet is the most massive one, the configuration is highly unstable with small values of $t_i$. These results allow to predict, for example, that a Jupiter size planet may destabilize more easily a pair of smaller planets (Super-Earths or Neptune size) in inner orbits rather than in outer orbits. 

The regression model also shows that the timespan of the chaotic period $t_c$ is longer when the initial separation between the planets is larger (larger values of $K$). In addition, if the outer planet is the less massive, $t_c$ is significantly shorter leading to a reduced period of 
close encounters between the planets. In all cases, the mass of the planet which is more easily thrown out of the system is the lowest among the three bodies and it happens in almost 70\% of P--P scattering events. 

It has been outlined that the complex network of mean motion and secular resonances among the planets (see for example \cite{charam2018})  may make inadequate the quadratic fit for some individual cases while, on average, the linear regression model works when used to outline the trends. This because the superposition of resonances may lead to a sudden decrease of $t_i$. However, this problem does not affect the fit to $t_c$ where the influence of resonances is negligible compared to the impulsive changes due to close gravitational approaches. 

It is debatable if a full machine learning approach might perform better by accounting somehow of the resonant web. However, it might improve the regression model with the risk of loosing control of the real important parameters of the system.

\section{Data availability}

The data underlying this article will be shared on reasonable request to the corresponding author.

\section{Acknowledgments} 
I thank an anonymous referee for helpful comments and suggestions. 

\bibliographystyle{aa.bst}

\bibliography{main}

\begin{thebibliography}{47}
\expandafter\ifx\csname natexlab\endcsname\relax\def\natexlab#1{#1}\fi

\bibitem[{{Chambers} {et~al.}(1996){Chambers}, {Wetherill}, \&
  {Boss}}]{chambers1996}
{Chambers}, J.~E., {Wetherill}, G.~W., \& {Boss}, A.~P. 1996, \icarus, 119, 261

\bibitem[{{Charalambous} {et~al.}(2018){Charalambous}, {Mart{\'\i}},
  {Beaug{\'e}}, \& {Ramos}}]{charam2018}
{Charalambous}, C., {Mart{\'\i}}, J.~G., {Beaug{\'e}}, C., \& {Ramos}, X.~S.
  2018, \mnras, 477, 1414

\bibitem[{{Chatterjee} {et~al.}(2008){Chatterjee}, {Ford}, {Matsumura}, \&
  {Rasio}}]{Chatterjee2008}
{Chatterjee}, S., {Ford}, E.~B., {Matsumura}, S., \& {Rasio}, F.~A. 2008, ApJ,
  686, 580

\bibitem[{{Davies} {et~al.}(2014){Davies}, {Adams}, {Armitage}, {Chambers},
  {Ford}, {Morbidelli}, {Raymond}, \& {Veras}}]{davies2014}
{Davies}, M.~B., {Adams}, F.~C., {Armitage}, P., {et~al.} 2014, in Protostars
  and Planets VI, ed. H.~{Beuther}, R.~S. {Klessen}, C.~P. {Dullemond}, \&
  T.~{Henning}, 787

\bibitem[{{Deienno} {et~al.}(2018){Deienno}, {Izidoro}, {Morbidelli}, {Gomes},
  {Nesvorn{\'y}}, \& {Raymond}}]{deienno2018}
{Deienno}, R., {Izidoro}, A., {Morbidelli}, A., {et~al.} 2018, \apj, 864, 50

\bibitem[{{Everhart}(1985)}]{radau1985}
{Everhart}, E. 1985, Astrophysics and Space Science Library, Vol. 115, {An
  efficient integrator that uses Gauss-Radau spacings}, ed. A.~{Carusi} \&
  G.~B. {Valsecchi}, 185

\bibitem[{{Fang} \& {Margot}(2013)}]{fang2013}
{Fang}, J. \& {Margot}, J.-L. 2013, \apj, 767, 115

\bibitem[{{Ghosh} \& {Chatterjee}(2023)}]{ghosh2023}
{Ghosh}, T. \& {Chatterjee}, S. 2023, \apj, 943, 8

\bibitem[{{Ghosh} \& {Chatterjee}(2024)}]{ghosh2024}
{Ghosh}, T. \& {Chatterjee}, S. 2024, \mnras, 527, 79

\bibitem[{{Gomes} {et~al.}(2004){Gomes}, {Morbidelli}, \&
  {Levison}}]{gomes2004}
{Gomes}, R.~S., {Morbidelli}, A., \& {Levison}, H.~F. 2004, \icarus, 170, 492

\bibitem[{{Lammers} {et~al.}(2023){Lammers}, {Hadden}, \&
  {Murray}}]{lammers2023}
{Lammers}, C., {Hadden}, S., \& {Murray}, N. 2023, \mnras, 525, L66

\bibitem[{{Laskar}(1993)}]{laskar1993}
{Laskar}, J. 1993, Celestial Mechanics and Dynamical Astronomy, 56, 191

\bibitem[{Laskar {et~al.}(1992)Laskar, Froeschl{\'e}, \& Celletti}]{laskar1992}
Laskar, J., Froeschl{\'e}, C., \& Celletti, A. 1992, Physica D: Nonlinear
  Phenomena, 56, 253

\bibitem[{{Laskar} \& {Petit}(2017)}]{laskar2017}
{Laskar}, J. \& {Petit}, A.~C. 2017, \aap, 605, A72

\bibitem[{{Lee} \& {Peale}(2002)}]{Leepeale2002}
{Lee}, M.~H. \& {Peale}, S.~J. 2002, \apj, 567, 596

\bibitem[{{Lin} \& {Ida}(1997)}]{lin1997}
{Lin}, D.~N.~C. \& {Ida}, S. 1997, \apj, 477, 781

\bibitem[{{Malmberg} \& {Davies}(2009)}]{mal2009}
{Malmberg}, D. \& {Davies}, M.~B. 2009, \mnras, 394, L26

\bibitem[{{Malmberg} {et~al.}(2011){Malmberg}, {Davies}, \& {Heggie}}]{mal2011}
{Malmberg}, D., {Davies}, M.~B., \& {Heggie}, D.~C. 2011, \mnras, 411, 859

\bibitem[{{Malmberg} {et~al.}(2007){Malmberg}, {de Angeli}, {Davies}, {Church},
  {Mackey}, \& {Wilkinson}}]{mal2007}
{Malmberg}, D., {de Angeli}, F., {Davies}, M.~B., {et~al.} 2007, \mnras, 378,
  1207

\bibitem[{{Marzari}(2014)}]{marzari2014}
{Marzari}, F. 2014, \mnras, 442, 1110

\bibitem[{{Marzari} \& {Nagasawa}(2019)}]{marzari_nagasawa2019}
{Marzari}, F. \& {Nagasawa}, M. 2019, \aap, 625, A121

\bibitem[{{Marzari} \& {Picogna}(2013)}]{marpi2013M}
{Marzari}, F. \& {Picogna}, G. 2013, \aap, 550, A64

\bibitem[{{Marzari} \& {Scholl}(2000)}]{marzari2000}
{Marzari}, F. \& {Scholl}, H. 2000, \apj, 543, 328

\bibitem[{{Marzari} \& {Weidenschilling}(2002)}]{marzari2002}
{Marzari}, F. \& {Weidenschilling}, S.~J. 2002, \icarus, 156, 570

\bibitem[{{Marzari} {et~al.}(2005){Marzari}, {Weidenschilling}, {Barbieri}, \&
  {Granata}}]{marza2005ApJ}
{Marzari}, F., {Weidenschilling}, S.~J., {Barbieri}, M., \& {Granata}, V. 2005,
  \apj, 618, 502

\bibitem[{{Masset} \& {Snellgrove}(2001)}]{masset2001}
{Masset}, F. \& {Snellgrove}, M. 2001, \mnras, 320, L55

\bibitem[{{Matsumura} {et~al.}(2008){Matsumura}, {Thommes}, {Chatterjee}, \&
  {Rasio}}]{matsu2008}
{Matsumura}, S., {Thommes}, E.~W., {Chatterjee}, S., \& {Rasio}, F.~A. 2008, in
  Astronomical Society of the Pacific Conference Series, Vol. 398, Extreme
  Solar Systems, ed. D.~{Fischer}, F.~A. {Rasio}, S.~E. {Thorsett}, \&
  A.~{Wolszczan}, 301

\bibitem[{{Moorhead} \& {Adams}(2005)}]{moor2005}
{Moorhead}, A.~V. \& {Adams}, F.~C. 2005, \icarus, 178, 517

\bibitem[{{Mustill} {et~al.}(2014){Mustill}, {Veras}, \&
  {Villaver}}]{mustill2014}
{Mustill}, A.~J., {Veras}, D., \& {Villaver}, E. 2014, \mnras, 437, 1404

\bibitem[{{Nagasawa} \& {Ida}(2011)}]{Nagasawa2011}
{Nagasawa}, M. \& {Ida}, S. 2011, ApJ, 742, 72

\bibitem[{{Obertas} {et~al.}(2023){Obertas}, {Tamayo}, \&
  {Murray}}]{obertas2023}
{Obertas}, A., {Tamayo}, D., \& {Murray}, N. 2023, \mnras, 526, 2118

\bibitem[{{Pedregosa} {et~al.}(2011){Pedregosa}, {Varoquaux}, {Gramfort},
  {Michel}, {Thirion}, {Grisel}, {Blondel}, {Prettenhofer}, {Weiss}, {Dubourg},
  {Vanderplas}, {Passos}, {Cournapeau}, {Brucher}, {Perrot}, \&
  {Duchesnay}}]{pedre2011}
{Pedregosa}, F., {Varoquaux}, G., {Gramfort}, A., {et~al.} 2011, Journal of
  Machine Learning Research, 12, 2825

\bibitem[{{Petrovich} {et~al.}(2014{\natexlab{a}}){Petrovich}, {Tremaine}, \&
  {Rafikov}}]{petrovich2014}
{Petrovich}, C., {Tremaine}, S., \& {Rafikov}, R. 2014{\natexlab{a}}, \apj,
  786, 101

\bibitem[{{Petrovich} {et~al.}(2014{\natexlab{b}}){Petrovich}, {Tremaine}, \&
  {Rafikov}}]{petrovich_rafikov2014}
{Petrovich}, C., {Tremaine}, S., \& {Rafikov}, R. 2014{\natexlab{b}}, \apj,
  786, 101

\bibitem[{{Pichierri} {et~al.}(2023){Pichierri}, {Bitsch}, \&
  {Lega}}]{pichi2023}
{Pichierri}, G., {Bitsch}, B., \& {Lega}, E. 2023, \aap, 670, A148

\bibitem[{{Picogna} \& {Marzari}(2014)}]{marpi2014}
{Picogna}, G. \& {Marzari}, F. 2014, \aap, 564, A28

\bibitem[{{Poon} {et~al.}(2020){Poon}, {Nelson}, {Jacobson}, \&
  {Morbidelli}}]{poon2020}
{Poon}, S. T.~S., {Nelson}, R.~P., {Jacobson}, S.~A., \& {Morbidelli}, A. 2020,
  \mnras, 491, 5595

\bibitem[{{Pu} \& {Wu}(2015)}]{pu2015}
{Pu}, B. \& {Wu}, Y. 2015, \apj, 807, 44

\bibitem[{{Rasio} \& {Ford}(1996)}]{rasioford1996}
{Rasio}, F.~A. \& {Ford}, E.~B. 1996, Science, 274, 954

\bibitem[{Raymond {et~al.}(2010)Raymond, Armitage, \& Gorelick}]{Raymond_2010}
Raymond, S.~N., Armitage, P.~J., \& Gorelick, N. 2010, The Astrophysical
  Journal, 711, 772

\bibitem[{{Raymond} {et~al.}(2008){Raymond}, {Barnes}, {Armitage}, \&
  {Gorelick}}]{raymond_barnes2008}
{Raymond}, S.~N., {Barnes}, R., {Armitage}, P.~J., \& {Gorelick}, N. 2008,
  \apjl, 687, L107

\bibitem[{{Raymond} {et~al.}(2009){Raymond}, {Barnes}, {Veras}, {Armitage},
  {Gorelick}, \& {Greenberg}}]{raymond2009}
{Raymond}, S.~N., {Barnes}, R., {Veras}, D., {et~al.} 2009, \apjl, 696, L98

\bibitem[{{Rice} {et~al.}(2018){Rice}, {Rasio}, \& {Steffen}}]{rice2018}
{Rice}, D.~R., {Rasio}, F.~A., \& {Steffen}, J.~H. 2018, \mnras, 481, 2205

\bibitem[{{Rice} {et~al.}(2016){Rice}, {Steffen}, \& {Rasio}}]{rice2016}
{Rice}, D.~R., {Steffen}, J.~H., \& {Rasio}, F.~A. 2016, in American
  Astronomical Society Meeting Abstracts, Vol. 227, American Astronomical
  Society Meeting Abstracts \#227, 138.28

\bibitem[{{Volk} \& {Gladman}(2015)}]{volk2015}
{Volk}, K. \& {Gladman}, B. 2015, \apjl, 806, L26

\bibitem[{{Weidenschilling} \& {Marzari}(1996)}]{weimar1996}
{Weidenschilling}, S.~J. \& {Marzari}, F. 1996, \nat, 384, 619

\bibitem[{{Zakamska} \& {Tremaine}(2004)}]{zaka2004}
{Zakamska}, N.~L. \& {Tremaine}, S. 2004, \aj, 128, 869

\end{thebibliography}

\appendix

\section {}
The fitting quadratic polynomial used in the linear regression model has the following form: 
\hfill \break

\begin{math}
log10(t) = b_0+b_1 m_1 + b_2 m_2 + b_3 m_3 + b_4 K + b_5   m_1^2 +b_6 m_2^2 +b_7 m_3^2 + b_8 K^2 + b_9 m_1 m_2 + b_{10} m_1 m_3 + b_{11} m_1 K +  b_{12} m_2 m_3 + b_{13}  m_2 K + b_{14} m_3 K
\end{math}
\hfill \break

The coefficients $b_i$ are given in the following table for both the regression model giving $t_i$ and $t_c$:

\begin{table}
\caption{Coefficients of the regression models}
\label{tab:example}
\begin{tabular}{lcc}
\hline
$b_i$ & $t_i$ & $t_c$ \\
\hline
$b_0$   & 2.913597 & 6.680226 \\
$b_1$   & -1.182263 & 0.562561 \\
$b_2$  & -2.560609 & -0.219297\\
$b_3$  &  1.662968 & -0.833863 \\
$b_4$  & -0.544556 & -0.163655 \\
$b_5$  & -0.629523 & -0.507074\\
$b_6$  & 0.226638 &  0.191720 \\
$b_7$  &  -0.466275 &  0.211043 \\
$b_8$  &  0.810959 &  -0.102188 \\
$b_9$  & -0.571602 &  -0.606726 \\
$b_{10}$  & 0.185247 & 0.800463 \\
$b_{11}$  &  1.148793 &  0.051381 \\
$b_{12}$  & -0.072094 & -0.339301 \\
$b_{13}$  & -0.708201  & 0.172606 \\
$b_{14}$  &  0.203866 & 0.024722 \\

\hline
\end{tabular}
\end{table}

While using these models it must be remembered that mutual Hill's sphere is defined iterativelly. In addition, the values of $t_i$
and $t_c$ must be scaled with the semi--major axis of the inner planet $a_1$, expressed in au, in the following way:
\hfill \break

\begin{math}
t_i (a_1) = t_{i0} \cdot \left ( a_1 / a_{10} \right ) ^{(3/2)}
\end{math}

\hfill \break
where $a_{10}$ is  equal to 5 au, the value adopted in all simulations, while $t_{i0}$ is the time derived from the above fit.

\label{lastpage}
\end{document}